\documentclass[a4paper,twocolumn,11pt]{quantumarticle}
\pdfoutput=1
\usepackage[utf8]{inputenc}
\usepackage[english]{babel}
\usepackage[T1]{fontenc}
\usepackage{amsmath}
\usepackage{hyperref}
\usepackage{amsfonts}
\usepackage{amssymb}
\usepackage[colorinlistoftodos, color=green!40, prependcaption]{todonotes}
\usepackage{amsthm}
\usepackage{mathtools}
\usepackage{physics}
\usepackage{xcolor}
\usepackage{graphicx}
\usepackage[left=23mm,right=13mm,top=35mm,columnsep=15pt]{geometry} 
\usepackage{adjustbox}
\usepackage{placeins}
\usepackage{lipsum}
\usepackage{csquotes}
\usepackage{enumitem}

\begin{document}

\title{Entanglement Signature of the Superradiant Quantum Phase Transition}

\author{Arthur Vesperini}
\affiliation{DSFTA, University of Siena, Via Roma 56, 53100 Siena, Italy}
\affiliation{QSTAR \& CNR - Istituto Nazionale di Ottica,    Largo Enrico Fermi 2, I-50125 Firenze, Italy}
\affiliation{INFN Sezione di Perugia, I-06123 Perugia, Italy}

\author{Matteo Cini}
\affiliation{Dipartimento di Fisica Università di Firenze, via G. Sansone 1, I-50019 Sesto Fiorentino, Italy}

\author{Roberto Franzosi}
\email[]{roberto.franzosi@unisi.it}
\affiliation{DSFTA, University of Siena, Via Roma 56, 53100 Siena, Italy}
\affiliation{QSTAR \& CNR - Istituto Nazionale di Ottica,    Largo Enrico Fermi 2, I-50125 Firenze, Italy}
\affiliation{INFN Sezione di Perugia, I-06123 Perugia, Italy}

\maketitle

\begin{abstract}
Entanglement and quantum correlations between atoms are not usually considered key ingredients of the superradiant phase transition. Here we consider the Tavis-Cummings model, a solvable system of two-levels atoms, coupled with a single-mode quantized electromagnetic field. This system undergoes a superradiant phase transition, even in a finite-size framework, accompanied by a spontaneous symmetry breaking, and an infinite sequence of energy level crossings. We find approximated expressions for the ground state, its energy, and the position of the level crossings, valid in the limit of a very large number of photons with respect to that of the atoms. In that same limit, we find that the number of photons scales quadratically with the coupling strength, and linearly with the system size, providing a new insight into the superradiance phenomenon.
Resorting to novel multipartite measures, we then demonstrate that this quantum phase transition is accompanied by a crossover in the quantum correlations and entanglement between the atoms (qubits). The latters therefore represent suited order parameters for this transition. Finally, we show that these properties of the quantum phase transition persist in the thermodynamic limit. 
\end{abstract}

\section{Introduction} \label{sec:outline}
The rapid experimental progress on quantum control is propelling the so-called second quantum revolution \cite{arute2019quantum,secondQR} forward, in which technologies based on the manipulation of single quantum systems, are being developed. In this light, quantum information plays a role of utmost relevance, for instance, in quantum cryptography, quantum computation, teleportation, the frequency standard improvement problem, quantum metrology and quantum artificial intelligence \cite{GUHNE20091}. Quantum correlation and quantum entanglement are fundamental resources in quantum information, whose potentialities are yet not fully unveiled.

Very recently, entanglement have been proposed as a fundamental resource for the realization of high-efficiency quantum batteries \cite{PhysRevLett.128.140501,PhysRevE.87.042123,Binder_2015}; more specifically, it has been shown that a quantum speedup, resulting in super-extensive power of quantum batteries, could theoretically be achieved by \textit{entangling operation} \cite{PhysRevE.87.042123,campaioli_enhancing_2017}. Many models adopted to describe systems realizing quantum batteries, are inspired by the Jaynes-Cummings, Tavis-Cummings, Rabi and Dicke models. It was shown in \cite{gyhm_quantum_2022} that the extensive advantage could not be achieved without global operations, that is, formally, non-local operators acting on the whole system at once; it is, however, not clear if systems such as the TC model fall into this category, as the interaction between the atomic degrees of freedom is somewhat indirect, \textit{mediated} by an electromagnetic cavity. It is therefore of great interest to assert if, the superradiance is accompanied by multipartite entanglement between the atoms, thus producing an \textit{effective} global entangling operation.

In a previous work \cite{PhysRevLett.115.180404} on the Rabi model, which describes the dynamics of a two-level atom coupled with a single-mode quantized field, it has been shown that this system undergoes a quantum phase transition (QPT), despite consisting of a finite number of degrees of freedom.
On the contrary, its multi-atom generalization, namely the Dicke model \cite{PhysRev.49.324,PhysRevLett.107.100401}, is known to undergo a QPT only in the thermodynamic limit.

The Jaynes–Cummings (JC) model was initially proposed in 1963 \cite{JCpaper} to describe the interaction of a two-level atom with an electromagnetic field. It consist of a simplification of the Rabi model (relying on the rotating-wave approximation), and is thus a fully solvable quantum model of a two-level atom (qubit) in interaction with a quantized single-mode field. The technical progress nowadays achieved, has made this system experimentally realizable \cite{PhysRevA.75.022312,Feng2015}, and this has renewed the interest to this class of models. Besides the applicative interest, the JC model is still considered an intriguing model by virtue of the many physical effects it exhibits, like, for instance, Rabi oscillations, collapses and revivals of Rabi oscillations and the superradiant phase transition \cite{Larson_2007,PhysRevA.101.053805,PhysRevA.101.043835,PhysRevA.98.021802}. Very recently, in Ref. \cite{PhysRevLett.117.123602} it has been shown that the JC model, despite the finite number of degrees of freedom of the system, also exhibits a QPT. 
In particular, this QPT is a second-order superradiant transition: in the broken-symmetry phase, the ground state corresponds to a photon condensate with a macroscopic photon occupation number. 

In this view, it is of crucial importance to carry out a study of the Tavis-Cummings (TC) model, which is the multi-atoms generalization of the JC model as well as an exactly solvable simplification of the Dicke model. The superradiant QPT has been demonstrated in this case, both in the resonant regime \cite{PhysRevLett.94.163601}, and in the off-resonant one \cite{castanos_coherent_2009}.

While entanglement between atoms has been addressed extensively in the bipartite case \cite{PhysRevLett.94.163601,dong_entanglement_2006,tessier_entanglement_2003,youssef_entanglement_2010}, it is not true, up to our best knowledge, of the multipartite case. In fact, only bipartite entanglement has been studied for multi-atoms systems \cite{PhysRevLett.94.163601}, whereas, as we will show, most of their entanglement content must be multipartite in nature.

In the present work, we thus tackle the study of the TC model, and obtain the following results:
\begin{itemize}
\item In the regime of a large number of photons with respect to the number of atoms, we derived an approximated expression for the eigenenergy and eigenvector of the ground state, and for the positions of the successive level crossings occurring in this QPT. 
\item In the same asymptotic regime, we show that the number of photons increases linearly with respect to the system size and quadratically with respect to the coupling strength. This turns out to be a good approximation soon after the transition point.
\item The QPT is accompanied by a crossover of the quantum correlations between the atoms, which converges, in the broken symmetry phase, to a finite value in the thermodynamic limit. We were able to perform this analysis thanks to a recently developed measure of quantum correlations, the Quantum Correlation Distance \cite{ScienticReports1-13}, which has the desirable property of possessing a closed-form very easy to compute. Consequently, this quantity plays the role of an order parameter for this QPT. 
\item The QPT is also accompanied by a crossover of the entanglement, i.e. a breaking of separability.  The degree of the said entanglement is not well captured by bipartite measures, such as the two-tangle. Leaning on the observation that quantum correlations maintain a high value, even in the thermodynamic limit, we conjecture that the entanglement per atom does not vanish either in this limit. 
\item Combining the previous observations lead to the hypothesis that the ground state of the TC model consists of genuinely multipartite entangled states, in the broken symmetry phase; the exact nature and magnitude of the entanglement of these states depends on an appropriate tuning of the parameters, that will be studied in more details in a future work.
\end{itemize} 

In this study, we limit ourselves to the zero-temperature regime, thus only investigating the ground state of the model and its dependence to the control parameters; thus, thermal decoherence effects are not taken into account.

This article is organized as follows: i) In the section {\it Methods} we summarize the relevant definitions concerning the quantum correlation distance (QCD) and entanglement distance (ED), which are measures for pure and mixed multi-partite states derived from some of us in former works \cite{j.aop.2019.167995, PhysRevA.101.042129,alireza,PhysRevA.105.052439,ScienticReports1-13,vesperini_arthur_geometry_2023}; ii) We introduce the TC model in section {\it Tavis-Cummings model}. iii) In section {\it Quantum phase transition} we address the superradiant phase transition undergone by the system at any finite size, showing that it corresponds to successive crossings, in a narrow region, of the subspaces of minimal energy. iv) Section {\it Quantum Correlations} shows our computations and results concerning the degree of quantum correlations between the two-level atoms, showing that the latter reaches a finite value in the thermodynamic limit. v) In Section {\it Quantum Entanglement}, 
we apply an entanglement criterion to various finite-sized system; then, computing the \textit{total two-tangle} in these systems, we apply the monogamy theorem \cite{osborne_general_2006} and recall the QCD previously calculated, to provide lower and upper bounds for the total amount of entanglement in the system. Our observations strongly suggest that the superradiant phase transition in this model is accompanied by a transition in the entanglement between the atoms, also in the thermodynamic limit.
{\it Conclusion} reports final comments.

\section{Methods} \label{sec:methods}
In the present paragraph, we summarize the definition of the QCD and the ED measures for pure and mixed states that some of us have derived in former works \cite{j.aop.2019.167995,PhysRevA.101.042129, alireza, PhysRevA.105.052439, ScienticReports1-13,vesperini_arthur_geometry_2023,GhofraneBelHadj:2023gie,vesperini_unveiling_2024,vesperini_entanglement_2024}.

Let ${\cal S}$ be the set of states on the tensor product of $M$- qubits Hilbert spaces ${\cal H } = {\cal H}^0 \otimes  \cdots \otimes {\cal H}^{M-1}$, i. e. the set of all positive operators with unit trace. 
We defined in Ref. \cite{ScienticReports1-13} the QCD per qubit for a state $\rho \in {\cal S}$ as
\begin{equation}
C(\rho) =\tr(\rho^2)- \frac{1}{M}\sum^{M-1}_{\mu=0} \lambda^\mu_{max} (\rho)  \, ,
\label{QCDrho}
\end{equation}
where $\mu=0,\ldots,M-1$ is a label for the qubits, $\lambda^\mu_{max} (\rho)$ are the maximal eigenvalues of the matrices $A^\mu(\rho)$ defined as follows
\begin{equation}
A^\mu_{ij}(\rho) = \tr [\rho \sigma^\mu_i
\rho\sigma^\mu_j ] \,  ,
\label{Amu}
\end{equation}
with $\sigma^\mu_j$ ($j=1,2,3$) the Pauli matrices acting on qubit $\mu$.
Note however that the QCD \eqref{QCDrho} scales as the purity. This is deemed an undesirable property for a measure of quantum correlations, as it implies that it may increase under the removal of an uncorrelated subsystem \cite{ozawaEntanglementMeasuresHilbertSchmidt2000a,pianiProblemGeometricDiscord2012a,changRemedyingLocalAncilla2013a,vesperini_arthur_geometry_2023} It can therefore be relevant to also consider a rescaled QCD. Following \cite{changRemedyingLocalAncilla2013a}, this issue can be solved by substituting the target density matrix $\rho$ with its square root $\sqrt{\rho}$, namely
\begin{equation}
\widetilde{C}(\rho):=C(\sqrt{\rho})\, .
\label{rescQCD}
\end{equation}
Another workaround for the purity-scaling problem, is to simply consider the quantity $C(\rho)/\tr(\rho)$. This somewhat more naive trick does not solve all of the problems raised by this scaling issue  \cite{ozawaEntanglementMeasuresHilbertSchmidt2000a,pianiProblemGeometricDiscord2012a,changRemedyingLocalAncilla2013a}.  It should however be noted that, in all the cases covered in the present work, it yields results sensibly identical to that of the rescaled QCD.

When computed on a pure state $\rho = |\psi\rangle\langle\psi|$, both Eqs. \eqref{QCDrho} and \eqref{rescQCD} reduce to a measure of entanglement, Entanglement Distance (ED) \cite{PhysRevA.101.042129,vesperini_arthur_geometry_2023,GhofraneBelHadj:2023gie}
\begin{equation}\label{EDpure}
E(|\psi\rangle)= 1 - \sum_{\mu=0}^{M-1} \sum_{j=1,2,3}\left|\langle\psi|\sigma_j^\mu|\psi\rangle\right|^2\, .
\end{equation} 
Yet, for mixed states, quantum correlations include, but \textit{are not limited to}, entanglement.
Given a state $\rho$, consider all of its possible decomposition $\{p_j,\rho_j\}$, such that
\begin{equation}
    \rho = \sum_j p_j \rho_j \, ,
\end{equation}
where $\sum_j p_j = 1$ and $\tr[\rho_j] = 1$.
A conventional method for building a mixed-state entanglement measure from any pure-state entanglement measure, is the convex roof construction \cite{vidal_2000}, namely
\begin{equation}\label{convex_roof}
E(\rho) = \min_{\{p_j,\rho_j\}} \sum_j p_j E(\rho_j)\, .
\end{equation}
The latter formula is not generally tractable, due to the exponential complexity of the minimization procedure. 
We have proposed \cite{ScienticReports1-13,vesperini_arthur_geometry_2023,GhofraneBelHadj:2023gie} a regularization procedure for $\rho$, that suppresses the quantum correlations unrelated to entanglement. By this procedure, the measure of quantum correlations \eqref{QCDrho} catches the true degree of entanglement of $\rho$. However, the implied procedure remains quite difficult to perform, and we were not able to apply it to the multipartite cases arising in this study.

\section{Tavis-Cummings model} \label{sec:jcm}
The TC model is an interesting generalization of the JC model, where $M$ two-level atoms (qubits) interacting with a single-mode of a quantized electromagnetic field are considered \cite{PhysRev.170.379}. The TC model, indeed, gives the opportunity to observe further non-classical effects, such as state squeezing and quantum state entanglement. These latter phenomena, in particular, are considered of primary importance, for instance, in pushing the performance of optical atomic clocks toward the Heisenberg limit \cite{PhysRevLett.104.073602}.

The TC Hamiltonian reads
    \begin{equation}
        H= \omega_c a^\dagger a + \omega_z S_3 - \frac{\lambda}{\sqrt{M}}(a^\dagger S_- + a S_+) \, ,
        \label{tcH}
    \end{equation}
where $a^\dagger$ and $a$ are the creation and annihilation operators of photons in the cavity and satisfying $[a,a^\dagger] = 1$. The total spin operators components $S_j$, $j=1,2,3$, satisfy the usual commutations $[S_i ,S_j ] = i \epsilon_{ijk} S_k$ and $S_\pm = S_1\pm i S_2$. Here and in the following we consider units in which $\hbar=1$. In terms of the Pauli matrices, we have $S_j = \sum^{M-1}_{\mu=0} \sigma^\mu_j/2$, where $\mu=0,\ldots,M-1$ runs on the index of the -distinguishable- atoms.
The model has three tuning parameters: the photon frequency $\omega_c$, the atomic energy 
splitting $\omega_z$, and the photon–atom coupling $\lambda$. 

The calculation of the eigenstates is achievable by writing down the Hamiltonian of the full system as a sum of two commuting terms.
Namely, $H= H_I + H_{II}$, where 
$H_{I}=\omega_c (a^\dagger a+ S_3)$, $H_{II}= \Delta S_3 - \lambda/\sqrt{M} (a^\dagger 
S_- + a S_+)$ and $\Delta=\omega_z-\omega_c$ is the detuning. Since $[H_I,H]=0$, $H_I$ is a conserved quantity. It results that, while the full Hilbert space is infinite-dimensional, the energy  spectrum can be organized in multiplets of -simultaneous- eigenstates for $H_I$ and $H_{II}$.

\section{Quantum phase transition} \label{sec:QPT}

The eigenvalues of the conserved quantity $H_I$ are $E^I_k = \omega_c (k-M/2)$ and their corresponding multiplicities $d^I_k = \min(k+1,M+1)$, with $k\in \mathbb{N}$ the total excitation number. Therefore, to determine the energy spectrum of the full Hamiltonian, we can diagonalize $H_{II}$ in each of the eigenspaces ${\cal H}^I_k$ of constant $k$ (of dimension $d^I_k$) associated with the $E^I_k$'s or, equivalently, with the $k$'s. Let us denote with $|n_{ph}, M_3\rangle$ the tensor product of a $n_{ph}$-photon Fock state and a normalized eigenstate of $S_3$ with eigenvalue $M_3$, $-M/2 \leq M_3\leq M/2$.
Note that the states $|n_{ph}, M_3\rangle$ are obviously also eigenstate of ${\bf S}^2$ with eigenvalues $M/2(M/2+1)$. The $k$-th eigenspace ${\cal H}^I_k$ is spanned by the vectors $|k,-M/2\rangle, \ldots, |0,k-M/2 \rangle$, for $k\leq M$ and $|k,-M/2\rangle, \ldots, |k-M,M/2 \rangle$, for $k>M$. 
Since we are investigating the properties of the ground state, we are solely interested in the lowest eigenvalue of $H$ in each ${\cal H}^I_k$, that we denote $E_k$.
The full Hamiltonian spectrum can be determined via numerical methods. 
Nevertheless, we can catch some hints about the mechanism at the base of the QPT, by investigating the properties of the lowest levels of ${\cal H}^I_k$, for $k=0,1$.

\begin{figure}[htb!]
    \centering
    {\includegraphics[width=1.\linewidth]{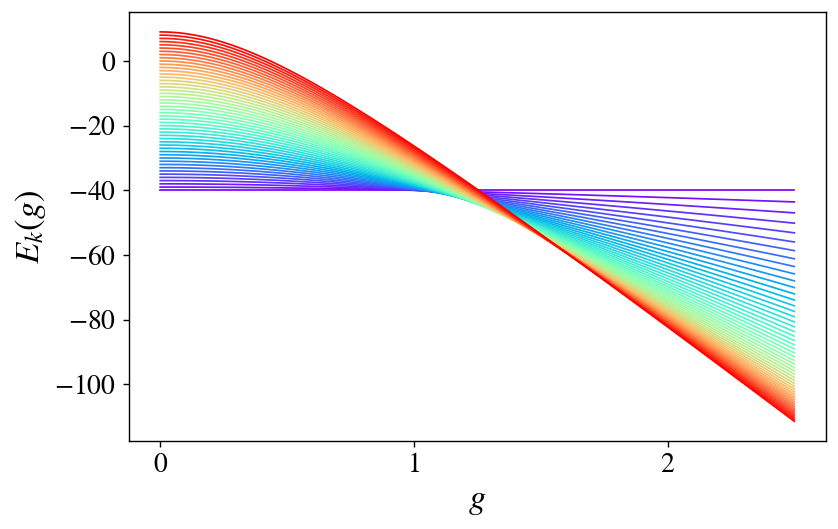}}
    \caption{Minima energy eigenvalues versus $g$, for the first fifty multiplets. The figure refers to a system of $M=8$ qubits and with $\eta=10$.}
    \label{crossing}
    \hspace{20.\parskip}
    \centering
    {\includegraphics[width=1.\linewidth]{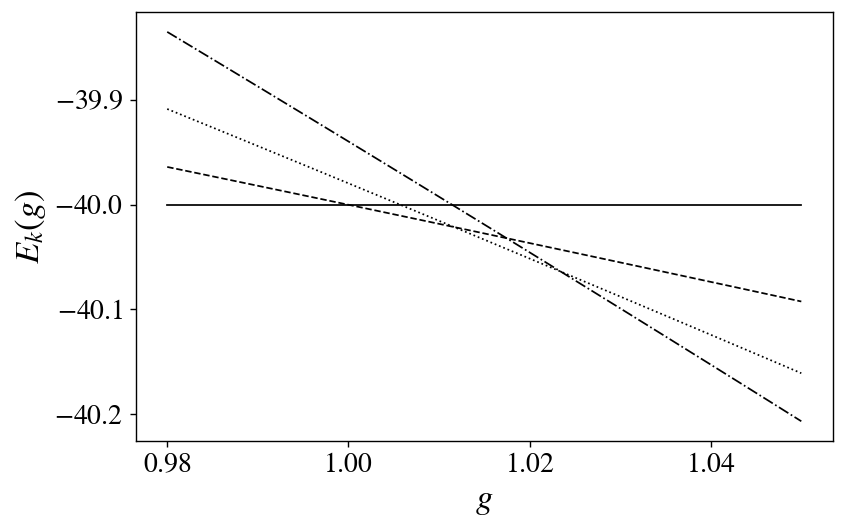}}
    \caption{The figure shows a zoom the energy levels $E_0$ (continuous line), $E_1$ (dashed line), $E_2$ (dotted line), $E_3$ (dot-dash line), $E_4$ (gray dot-dash line) versus $g$. The parameters are the same as for figure \ref{crossing}.}
    \label{crossingzoom}
\end{figure}

The vacuum state $|E_0 (g)\rangle=|0,-M/2 \rangle$ is the eigenstate in ${\cal H}^I_0$ with eigenvalue $E_0(g) = -\omega_z M/2$. The minimum energy level of $H$ in ${\cal H}^I_1$
reads
\begin{multline}
E_1(g) = \dfrac{\omega_c}{2} +\dfrac{\omega_z}{2}\left(1- M\right)
 \\
-\frac{\omega_z}{2} \sqrt{\left(1-\dfrac{1}{\eta}\right)^2 + \dfrac{4g^2}{\eta}} \, ,
    \label{ievl1}
\end{multline}
where we have introduced the dimensionless parameters $\eta=\omega_z/\omega_c$ and
$g=\lambda/\sqrt{\omega_c \omega_z}$. For the sake of simplicity, we will drop the dependences on the parameter $\eta$, that we fixed to $\eta=10$ throughout this work. 
The corresponding eigenstate is
\begin{equation}
|E_1 (g) \rangle = s |0,1-M/2\rangle - c |1,-M/2\rangle \, ,
    \label{eist1}
\end{equation}
where $s=\sin(\beta/2)$ and $c=\cos(\beta/2)$, with
\begin{equation}
    \beta = \arccos \left[\frac{1-\eta^{-1}}{\sqrt{(1-\eta^{-1})^2 +4g^2\eta^{-1}}}\right] \, .
    \label{beta1}
\end{equation} 
For $g<1$, the ground state of the full Hamiltonian \eqref{tcH} is $|E_0 (g)\rangle$.
Yet, at $g=1$ level $E_0(g)$ crosses level $E_1(g)$, which then becomes the ground state energy of the full Hamiltonian. This first level crossing is followed by further crossings between minimum energy levels of successive eigenspaces ${\cal H}^I_k$. Formally, \begin{equation*}
\begin{split}
\forall k,\, \exists!\, g_k\text{ s.t. }&\forall g<g_k,\,E_k(g)<E_{k+1}(g),\\ \text{ and }&\forall g>g_k,\,E_k(g)>E_{k+1}(g).
\end{split}
\end{equation*} In other words, as $g$ increases, the ground state of the full Hamiltonian is successively found in eigenspaces ${\cal H}^I_k$ of larger $k$.
Figure \ref{crossing} shows the crossings of the energy levels $E_k$, for $k=0,\ldots,50$. Figure \ref{crossingzoom} is a zoom, showing the crossing of the levels $E_0,E_1,E_2,E_3,E_4$. 

At $g=1$, the system undergoes a quantum phase transition under the form of a spontaneous symmetry breaking. 
In fact, the Hamiltonian \eqref{tcH} is invariant under the continuous $U(1)$ symmetry group associated to the unitary operators $e^{iT\phi}$, where $T=H_{II}+\mathbb{I} \Delta M/2$ and $\phi \in \mathbb{R}$, since $[H,T]=0$.
Yet, for $g<1$ the ground state is invariant under the action of such operators, since $T|E_0\rangle =0$, on the contrary, for $g>1$ the ground state is no longer invariant under the same symmetry, bringing the system to a SSB.

In the asymptotic limit $k \gg M$, the following approximation for the full Hamiltonian holds
\begin{multline}
H \approx \omega_c \left(k-\dfrac{M}{2}\right) \mathbb{I} +\omega_z \left(1-\dfrac{1}{\eta}\right) J_3  \\ 
-2 \omega_z g  \sqrt{\dfrac{k}{\eta M}} J_1 \, ,
    \label{Happrox}
\end{multline}
where the operators $J_j$, for $j=1,2,3$, are the usual angular momenta operators. 
We introduce the usual basis of eigenstates for ${\bf J}^2$ and $J_3$ derived from the basis vectors of ${\cal H }^I_k$, according to the following mapping 
\begin{equation}
|J=M/2,\,J_3 =M_3\rangle:=|k - M/2-M_3,\,M_3\rangle\, , 
\end{equation}
for $M_3=-M/2,\ldots,M/2$. 
The minimum energy level of the approximated Hamiltonian \eqref{Happrox} in each eigenspace ${\cal H}^I_k$ writes
\begin{multline}
\tilde{E}_k(g) = \omega_c  \left(k - \dfrac{M}{2} \right) \\
- \omega_z  \dfrac{M}{2} \sqrt{\left(1-\dfrac{1}{\eta}\right)^2 + \dfrac{4g^2k}{\eta M}} \, ,
    \label{ievlk}
\end{multline}
and the corresponding eigenvector is
\begin{equation}
|\tilde{E}_k (g)\rangle = \sum^{M}_{n=0} \binom{M}{n}^{\frac{1}{2}}
c_k^{M-n}
s_k^{n} |k-n,n-M/2 \rangle\, ,
    \label{eistk}
\end{equation}
where $c_k = \cos (\beta_k/2)$, $s_k = \sin (\beta_k/2)$ and
\begin{equation}
    \beta_k = \arccos{ \left[\frac{(1-\eta^{-1})}{\sqrt{(1-\eta^{-1})^2 + 4 g^2 \eta^{-1}k/M}} \right]} \, .
    \label{betak}
\end{equation}
\begin{figure}[htb!]
    \centering
    {\includegraphics[width=1.\linewidth]{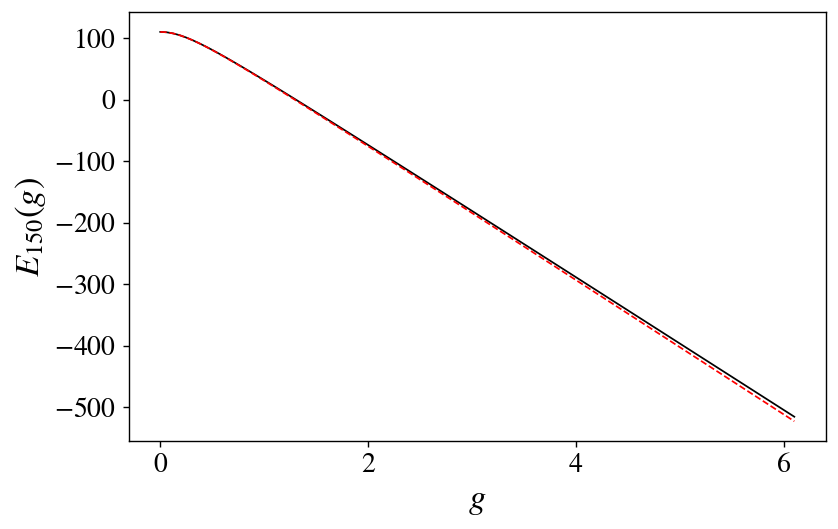}}
    \caption{The figure compares the plots of the energy level $E_{150}$ versus $g$, corresponding to the minimum energy eigenvalue of the eigenspace ${\cal H}^I_k$, $k=150$. Here we have considered a system with $M=8$ qubits and with $\eta=10$. In continuous black line, we report $E_{150}(g)$  derived by numeric diagonalization of the full Hamiltonian \eqref{tcH} and in dashed red line the approximated level given in Eq. \eqref{ievlk}. The agreement between the two curves is very good.}
    \label{e50}
\end{figure}
In Fig. \ref{e50} we compare the plots of $E_{150}(g)$, derived by direct numeric diagonalization of the full Hamiltonian \eqref{tcH}, and of $\tilde{E}_{150}(g)$, computed from Eq. \eqref{ievlk}. The agreement between the two plots is very good on a wide range of values for $g$.
Furthermore, after Eq. \eqref{ievlk}, it is easy to prove that the crossing between $\tilde{E}_k(g)$ and $\tilde{E}_{k+1}(g)$ occurs at the point
\begin{equation}\label{approx_gk}
\tilde{g}_k \approx \left\{ \dfrac{2 k}{ \eta M }
 \left[
1+\sqrt{1+\left(\frac{M(\eta-1)}{2k}\right)^2
}
\right]
\right\}^{\frac{1}{2}} \, .
\end{equation}
Remarkably, in the limit of strong spin energy separation, that is $\eta \to \infty$, all the crossing-level points merge at the quantum phase transition point $g =1$.

Note that, from Eq. \eqref{approx_gk}, the distance between two successive level crossings scales as
\begin{equation}\label{dgk}
d\tilde{g}_k = \left(\tilde{g}_{k+1} -\tilde{g}_k\right) \sim o\left( \sqrt{\frac{1}{k\eta M}} \right)\, .
\end{equation}
This suggests that, also in limit $M\to\infty$, all the level crossings merge at $g=1$. 

Remarkably, the relation \eqref{approx_gk} can be reversed to obtain the excitation number $\tilde{k}^*$ associated to the ground state, 
\begin{equation}\label{approx_kstar}
\tilde{k}^*\approx \frac{g^2\eta M}{4}\, ,
\end{equation}
valid in the regime $k\gg M$. Note that, in this regime, $n_{ph}\approx k$, and $k^*$ may thus be considered as the number of photons in the system. \\

\begin{figure}[htb!]
    \centering
    {\includegraphics[width=1.\linewidth]{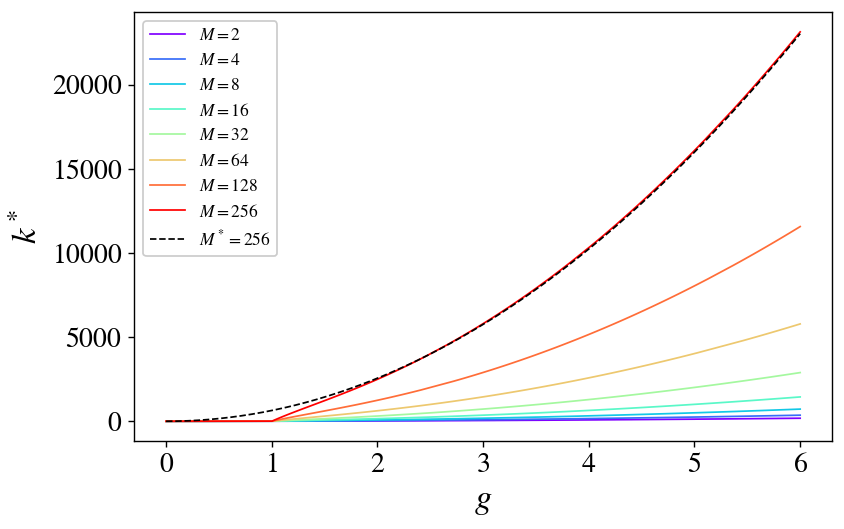}}
    \caption{The figure reports the excitation number $k^*$ of the ground state, as a function of the coupling $g$, derived by numerical calculations, in the cases $M=2,3,4,5,6,7,8,9$. Line $M^*=9$ shows the asymptotic prediction \eqref{approx_kstar}, valid in the limit $k \gg M$.}
    \label{kstar_TC}
    \hspace{20.\parskip}
    \centering
    {\includegraphics[width=1.\linewidth]{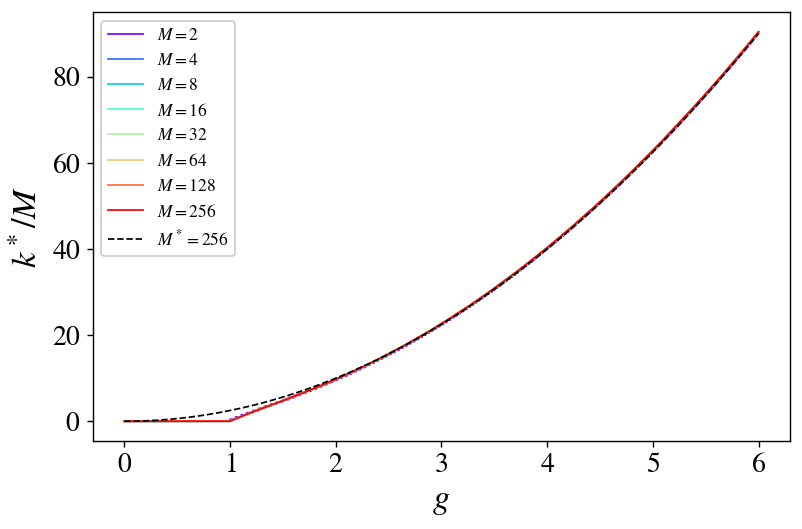}}
    \caption{The figure reports the excitation number per qubit $k^*/M$ of the ground state, as a function of the coupling $g$, derived by numerical calculations. All the curves, corresponding to the cases $M=2,3,4,5,6,7,8,9$, are completely overlapping, and represented by the single red curve. The dotted line shows the asymptotic prediction drawn from \eqref{approx_kstar}, valid in the limit $k \gg M$.}
    \label{kstarM_TC}
\end{figure}

In Figure \ref{kstar_TC}, we show the behaviour of the excitation number $k^*$ of the ground state as a function of $g$. The curves $k^*/M$, shown in figure \ref{kstarM_TC}, all collapse onto the same curve, and agreement with the analytic approximation $k^*/M\approx \eta g^2 /4$ (see Eq. \eqref{approx_kstar}) is very good, especially when $g\gtrsim2$. 
It results that $k^*$ is linear in $M$ and quadratic in $g$, up to very small corrections. 
This suggests that, in the thermodynamic limit, $k^*$ diverges for all $g\gtrsim1$, and all the level crossings thus collapse at $g\approx 1$, accounting for the phenomenon of superradiance.

Note that, as predicted in Eq. \eqref{dgk}, the precision on $g$ required to capture all the level-crossings is of order $o\left(M k\right)^{-1/2}$. The complexity of the computations is all the more increased, considering higher $M$.

\section{Quantum correlations} \label{sec:quantum correlation}

In this section, our main focus lies in examining how the system responds to changes in the coupling parameter $g$. We thus set $\eta$ and $\omega_c$ to reasonable fixed values, namely $\eta=10$ and $\omega_c=1$, as we did in the previous figures. 
The QCD between the atoms is computed by applying definition \eqref{QCDrho} to the atomic density matrix $\rho_s (g)$. The latter is derived from the density matrix $\rho(g)$ describing the ground state of the full system, by tracing out the photonic part of the system, namely $\rho_s (g)=\tr_{ph} [\rho (g)]$.

The $g$-dependent ground state of the full system write
\begin{equation}
\begin{split}
|GS (g) \rangle & =  |E_k(g) \rangle\, , \\
\text{where  } E_k (g) & = \min_n \{ E_n (g) \}\, .
\end{split}
\end{equation}
In the general case, we compute the QCD by numeric calculations. Nevertheless, we present hereafter the few analytically tractable closed-form results, namely the cases $g<1$, $g\gtrsim 1$ and $k\gg M$. Doing so, we aim at a better understanding of the behaviour of the QCD in the $g$-parametric space. Furthermore, it provides a well-suited test for the validity of the approximated Hamiltonian \eqref{Happrox}. 

In our calculation, we will resort to the Dicke states $|D^{M}_n\rangle$ of $M$-qubits.
The eigenstate $|S=\frac{M}{2}, M_3 = n - \frac{M}{2}\rangle=|D^M_n\rangle$ of $S_3$, of eigenvalue $M_3=n - \frac{M}{2}$, is in fact a degenerate state of $n$ excited qubits. Thus, it 
equates the Dicke state
 \begin{equation}
|D^{M}_n\rangle =\binom{M}{n}^{-\frac{1}{2}}
\sum_j P_j\left\{|e\rangle^{\otimes n} \otimes |g\rangle^{\otimes M-n}\right\} \, ,
\label{dickes}
\end{equation}
where we denote with $\sum_j P_j$ the sum over all the possible permutations and, for $\mu=0,\ldots,M-1$, $|e\rangle^\mu$ and $|g\rangle^\mu$ are the eigenstates of $\sigma^\mu_3$ with eigenvalues $+1$ and $-1$, respectively. It can easily be verified that $S_3|D^{M}_{n}\rangle=\left(n-\frac{M}{2}\right)|D^{M}_{n}\rangle$.

The atomic part of the ground state can thus always be expressed as
\begin{equation}\label{general_D_mix}
\rho_s (g) = \sum_{n=0}^M p_n | D^M_{n} \rangle\langle D^M_{n} | \, ,
\end{equation}
namely, a mixture of Dicke states. Yet, the Dicke states \eqref{dickes} are clearly symmetric under any permutation of the qubits, and so is any mixture of Dicke states. The quantities of interest can thus always be calculated considering a single arbitrary qubit $\mu$.

As a preliminary to the computation of the QCD in various cases, the following will prove necessary
\begin{equation}
\begin{split}
\langle D^M_{n} | \sigma^\mu_1 | D^M_{n^\prime} \rangle &= 
\sqrt{\left(\dfrac{M-n}{M} \right) \left( \dfrac{n+1}{M} \right)} \delta_{n ,n^\prime-1} \\
&+ \sqrt{\left(\dfrac{(M-n+1)}{M} \right) \dfrac{n}{M}} \delta_{n ,n^\prime+1}
 \, , \\
\langle D^M_{n} | \sigma^\mu_2 | D^M_{n^\prime} \rangle &= i\sqrt{\left(\dfrac{M-n}{M} \right) \left( \dfrac{n+1}{M} \right)} \delta_{n ,n^\prime-1} \\
&- i \sqrt{\left(\dfrac{(M-n+1)}{M} \right) \dfrac{n}{M}} \delta_{n ,n^\prime+1}
 \, , \\
\langle D^M_{n} | \sigma^\mu_3 | D^M_{n^\prime} \rangle &= 
\left(
\dfrac{2n}{M}-1
\right) \delta_{n, n^\prime} 
\, .\label{DMn_sig_DMm}
\end{split}
\end{equation}
From the latter identities, it is indeed possible to derive the eigenvalues $\lambda^\mu_{i} (\rho_s)$ ($i=1,2,3$) of the matrices $A^\mu(\rho_s)$, as defined in Eq. \eqref{Amu}.

After some calculation, we obtain the following formulas, valid in the most general case
\begin{equation}\label{eivals_general}
\begin{split}
\lambda^\mu_{1(2)}(\rho_s)
= &2\sum^{M-1}_{n=0} p_n p_{n+1}  \left(\frac{M-n}{M}\right)\left(\frac{n+1}{M}\right)\\
\lambda^\mu_{3}(\rho_s)
= &\sum^{M}_{n=0} p_n^2 \left(\dfrac{2n}{M}-1\right)^2 \, ,
\end{split}
\end{equation}
where $\lambda^\mu_{1(2)}(\rho_s)  = \lambda^\mu_{1} (\rho_s)= \lambda^\mu_{2} (\rho_s)$.
\paragraph{\textbf{Case} ${\bf g<1}$}
In this case, the full system ground state is $|E_0(g)\rangle$ and $\rho_s (g) = \tr_{ph}[ |E_0(g)\rangle \langle E_0(g)|] = |D^M_0\rangle \langle D^M_0|$. We hence have $p_n=\delta_{n,0}$. Remark that $\rho^2_s=\rho_s$, i.e. the ground state is \textit{pure}.
From Eq. \eqref{eivals_general} we get
\begin{equation}
C\big(\rho_s(g<1)\big) = 0.
\end{equation}

\paragraph{\textbf{Case} ${\bf g\gtrsim 1}$}
$|GS(g)\rangle=|E_1(g)\rangle$ reported in Eq. \eqref{eist1} and we have 
\begin{equation}
\rho_s (g\gtrsim1) = s^2 |D^M_{1}\rangle \langle D^M_{1}| + c^2 |D^M_0\rangle \langle D^M_0| \, ,
\label{rhos}    
\end{equation}
where $s=\sin(\beta/2)$, $c=\cos(\beta/2)$ and $\beta$ is given in Eq. \eqref{beta1}. 
Remarkably, the fact that the partial trace operation results in a mixed state, highlights that the photonic and atomic parts of the system are entangled in this case.

By direct calculation we derive the purity $\tr (\rho_s^2) = s^4+c^4$. Furthermore, inserting $p_n = s^2\delta_{n,1} + c^2\delta_{n,0}$ in Eq. \eqref{eivals_general}, we obtain 
\begin{equation}
\lambda^\mu_{max} \big(\rho_s(g\gtrsim1)\big) = s^4+c^4 - 4 \frac{s^4}{M}(1-\frac{1}{M}) \, ,
\end{equation}
thus
\begin{equation}
C\big(\rho_s(g\gtrsim 1)\big) = \dfrac{4 s^4}{M} \left(1-\dfrac{1}{M} \right) \, .
\end{equation}

\paragraph{\textbf{Case} ${\bf k\gg M}$}
We now consider the approximated ground states of Eq. \eqref{eistk}, $|GS(g)\rangle \approx |\tilde{E}_k(g)\rangle$. 
Let us denote $\tilde{\rho}_s$ the corresponding atomic density matrix. Its probabilistic weights write
\begin{equation}\label{p_n_asymp}
p_n(k\gg M):=\alpha_{kn}=\binom{M}{n} c_k^{2(M-n)} s_k^{2n}.
\end{equation}
Once again, the mixedness of this state indicates a strong entanglement between the photonic and atomic parts of the system.
By direct calculation we get the purity
\begin{equation}
\tr (\tilde{\rho}_s^2)= \sum^{M}_{n=0} \alpha_{kn}^2 \, ,
\end{equation}
and, inserting $p_n=\alpha_{kn}$ in Eq. \eqref{eivals_general}, the two eigenvalues of $A^\mu$
\begin{equation}\label{eivals_approx}
\begin{split}
\lambda^\mu_{1(2)} \big(\tilde{\rho}_s\big) = &2\sum^{M}_{n=0} \alpha_{kn}^2\left(\frac{s_k}{c_k}\right)^2\left(\dfrac{M-n}{M}\right)^2   \\
\lambda^\mu_{3} \big(\tilde{\rho}_s\big)= &\sum^{M}_{n=0} \alpha_{kn}^2 \left(2\dfrac{n}{M}-1\right)^2\, .
\end{split}
\end{equation}
Finally, we get 
\begin{equation}\label{qcdapprox}
\begin{split}
C(\tilde{\rho}_s(g)) =\min\left\{ \sum^{M}_{n=0}  \alpha_{kn}^2\beta_n,\, \sum^{M}_{n=0}\alpha_{kn}^2\gamma_n\right\}\, ,
\end{split}
\end{equation}
with
\begin{equation}
\begin{split}
&\beta_n= 4\dfrac{n}{M}\left(\frac{M-n}{M}  \right),\\
&\gamma_n= \left( 1 - 2 \left(\frac{s_k}{c_k}\right)^2\left(\frac{M-n}{M}\right)^2 \right)\, .
\end{split}
\end{equation}

\begin{figure}[htb!]
    \centering
    {\includegraphics[width=1.\linewidth]{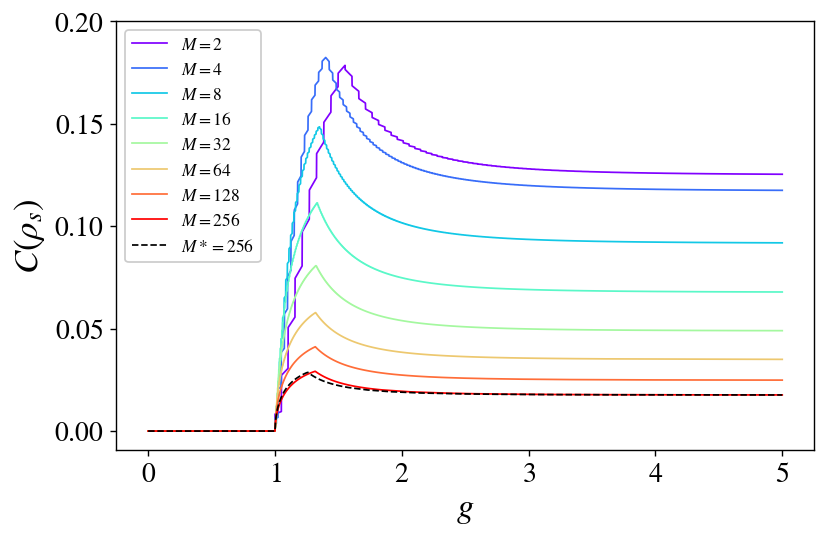}}
    \caption{The figure plots the QCD per qubits $C(\rho_s)$ versus $g$, derived by numerical calculations. The lines refer to the cases of a system with $M=2,3,4,5,6,7,8,9$ qubits. Line $M^*=9$ shows the asymptotic prediction \eqref{qcdapprox}, valid in the limit $k \gg M$.}
    \label{QCD_TC}
\end{figure}
Figure \ref{QCD_TC} reports the QCD per qubit $C(\rho_s)$ as a function of $g$, achieved by numerical calculations. The lines refer to the cases $M=2,3,4,5,6,7,8,9$ and the asymptotic predictions, valid in the limit $k \gg M$, for the case $M^*=9$ (see the legend). The agreement between the numerical results and the analytical prediction valid in the limit $k \gg M$ is very good in the strong coupling regime. 

Note that, in the region $1<g<2$, the QCD displays an inflexion point in the form of a peaked discontinuity. The latter corresponds to a crossover between the two eigenvalues presented in Eq. \eqref{eivals_general}.

Here, it is evident that the QCD is decreasing in average, as a function of $M$, even around the transition, at $g\gtrsim1$. We however claim that this behaviour should be considered an artefact, due to the purity-scaling of QCD. 
\begin{figure}[htb!]
    \centering
    {\includegraphics[width=1.\linewidth]{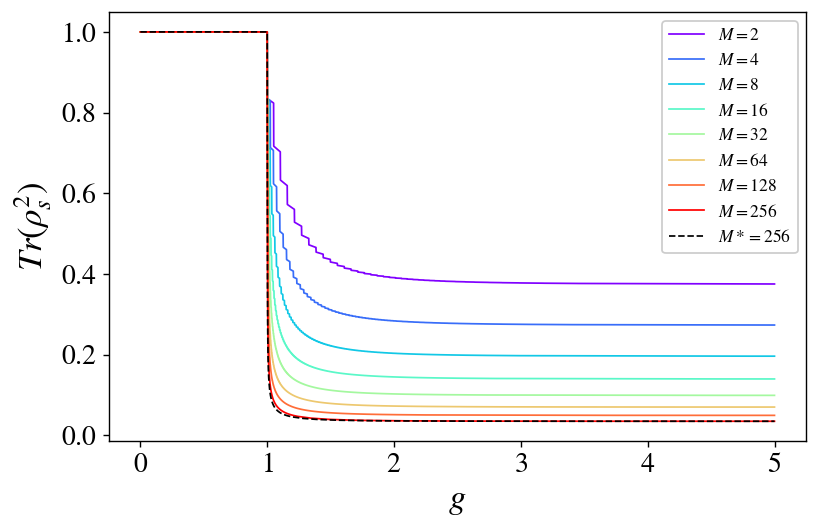}}
    \caption{The figure reports the purity $\tr(\rho_s^2)$ as a function of the coupling $g$, derived by numerical calculations, in the cases $M=2,3,4,5,6,7,8,9$. }
    \label{purity_TC}
\end{figure}
We indeed verified, as shown in Figure \ref{purity_TC}, that the purity $\tr(\rho_s^2)$ is a decreasing function of $g$ and a decreasing function of $M$.\\

Motivated by this observation, we thus deem especially useful in this case to also consider the rescaled measure \eqref{rescQCD}. 

The Dicke states form an orthonormal basis, so $\sqrt{\rho_s}$ is straightforwardly obtained by substituting, in Eqs. \eqref{general_D_mix} and \eqref{eivals_general}, the probability weights $p_n$ (i.e. the eigenvalues of $\rho_s$) with their square root $\sqrt{p_n}$. It results, in the case $g\gtrsim 1$,
\begin{equation}
\widetilde{C}(\rho_s(g)) = \frac{4 s^2}{M} \left(\frac{1}{2}-\dfrac{1}{M} \right)\, ,
\end{equation}
and, in the case $k\gg M$,
\begin{equation}\label{qcdrescapprox}
\begin{split}
\widetilde{C}(\tilde{\rho}_s(g)) =\min\left\{ \sum^{M}_{n=0} \alpha_{kn} \beta_n,\;\sum^{M}_{n=0} \alpha_{kn} \gamma_n' \right\}\, ,
\end{split}
\end{equation}
with
\begin{equation}
\gamma'_n = 1 - 2 \frac{s_k}{c_k}\left(\frac{M-n}{M}\right)^{\frac{3}{2}}\left(\frac{n+1}{M}\right)^{\frac{1}{2}}
\end{equation}

\begin{figure}[htb!]
    \centering
    {\includegraphics[width=1.\linewidth]{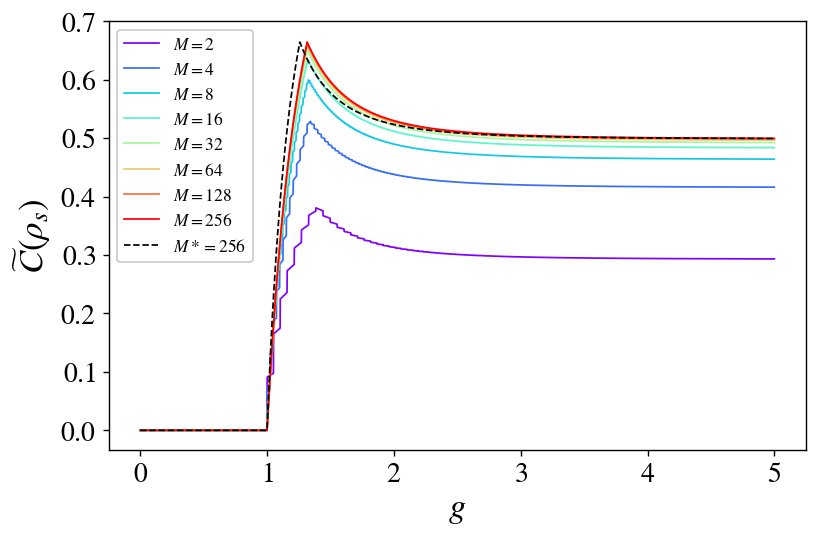}}
    \caption{The figure reports the rescaled QCD per qubit $\widetilde{C}(\rho_s)$ as a function of the coupling $g$, derived by numerical calculations, in the cases $M=2,3,4,5,6,7,8,9$. Line $M^*=9$ shows the asymptotic prediction \eqref{qcdapprox}, valid in the limit $k \gg M$.}
    \label{QCDresc_TC}
\end{figure}
Figure \ref{QCDresc_TC} reports the rescaled QCD per qubit $\widetilde{C}(\rho_s(g))$ as a function of $g$, achieved by numerical calculations. Once again, the agreement between the numerical results and the analytical prediction valid in the limit $k \gg M$ is very good in the strong coupling regime. 
The rescaled QCD clearly converges, with increasing $M$, toward a finite value both at $g\gtrsim 1$ (where it approaches a value greater than $0.6$) and for $g\gg1$ (where it approaches $0.5$). 

Most interestingly, when the rescaled QCD is considered (thus we substitute $p_n$ with $\sqrt{p_n}$), the third eigenvalue of Eq. \eqref{eivals_general} becomes
\begin{equation}
\lambda_3^\mu(\sqrt{\rho_s} )=\sum_{n=0}^M p_n \left|\langle D^M_n|\sigma_3|D^M_n\rangle\right|^2\,.
\end{equation}
In the $g$-region previous to the peak, this eigenvalue is the largest. This yields
\begin{equation}\label{QCD_eq_roof}
\Tilde{C}(\rho_s (g<g_M))=\sum_{n=0}^M p_n E\left(|D^M_n\rangle\right)\, ,
\end{equation}
where $g_M$ is the $M$-dependent position of the peak and $E$ is the ED, as defined in Eq. \eqref{EDpure}. .
Thus, for $g<g_M$, the rescaled QCD equates the average entanglement of the Dicke states entering in the statistical mixture $\rho_s$. On the contrary, for $g>g_M$, the first eigenvalues of Eq. \eqref{eivals_general} become larger than the third one. The QCD then becomes smaller than this average. Quantum correlations are hence diminished in this latter case by the \textit{interaction} between the Dicke states entering the mixture. One can grasps an intuition of this fact by considering the shape of the eigenvalues in Eq. \eqref{eivals_general}. 
Ultimately, this crossover can be attributed to the fact that the distribution $p_n(g)$ becomes more evenly spread as $g$ increases, while almost only the first excitation numbers take finite values for $g<g_M$.

As anticipated, quantum correlations between atoms is null for $g\leq1$, and take finite values for $g>1$, thus behaving as an order parameter marking the superradiant transition.

\section{Quantum entanglement} \label{sec:quantumentanglement}

Quantifying the entanglement of an arbitrary state is a notoriously challenging task. Unfortunately, Eq. \eqref{convex_roof} is not a priori tractable for general Dicke mixtures. We however have at our disposal a few tools to investigate the amount and nature of the entanglement among the assembly of atoms.

In Ref. \cite{yuSeparabilityMixtureDicke2016}, a separability criterion for mixtures of Dicke states was proposed. Given a state of the form $\rho=\sum_j p_j|D^M_j\rangle\langle D^M_j|$, it relies on the positive semi-definiteness of Hankel matrices formed with the weights $p_j$; an equivalent formulation of the same criterion involves the positive semi-definiteness of $\rho^\Gamma$, where $\Gamma$ stands for the partial transpose of $\left \lfloor{\frac{N}{2}}\right \rfloor $ subsystems.\\
We applied both of these methods on the atomic reduced ground states $\rho_s$ and found that, for each considered $M$, it is entangled $\forall g>1$, and separable otherwise; this result is independent from the value of the other parameters, $\eta$ and $\omega_c$.
Unfortunately, this criterion, as effective as it may be, lacks a quantitative aspect. Hence, though $\rho_s(g>1)$ is entangled for all the values of $M$ we considered, the \textit{intensive} degree of entanglement could be tending to zero as we approach the thermodynamic limit.

For lack of a general computationally affordable measure of entanglement, we can define bounds for $E(\rho_s)$ as defined in Eq \eqref{convex_roof}. 
Let us first define the \textit{total two-tangle per qubit} \cite{osborne_general_2006}
\begin{equation}
\tau_{tot}(\rho_s)=\frac{1}{M}\sum_{(\mu\nu)}{\cal C}_{\mu\nu}^2(\rho_s)\, ,
\end{equation}
where ${\cal C}_{\mu\nu}(\rho):={\cal C}\left(\rho_{\mu\nu}\right)$ is the concurrence computed on the two-qubit reduced state $\rho_{\mu\nu}=\tr_{\bar{\mu\nu}}(\rho)$, obtained by tracing out the complement of the pair $(\mu,\nu)$ in the set of all the qubits. The sum is running over all the pairs of distinct qubits. 

The factor $\frac{1}{M}$ insures that this quantity is intensive, and thus comparable with the other quantities presented in this work. This is for instance in contrast with the convention chosen in Ref. \cite{buzek_instability_2005}.
It is a well-known result that, in virtue of the \textit{monogamy of entanglement}, this quantity stems as a lower bound for the total amount of entanglement in the system \cite{osborne_general_2006,coffman_distributed_2000}. It is easy to verify that the measures of the total entanglement employed in the latter references are in fact equivalent to Eq. \eqref{convex_roof}. The monogamy theorem is thus valid in this context.

We can also consider the valuable information provided the QCD and rescaled QCD in this context. In fact, the latter provides an upper bound for the total amount of entanglement in the system, as measured by Eq. \eqref{convex_roof}. Furthermore, as emphasized in the previous section, in the region $g<g_M$, the QCD values as shown in Eq. \eqref{QCD_eq_roof}, which constitutes an obvious roof for Eq. \eqref{convex_roof}.

\begin{figure}[htb!]
    \centering
    {\includegraphics[width=1.\linewidth]{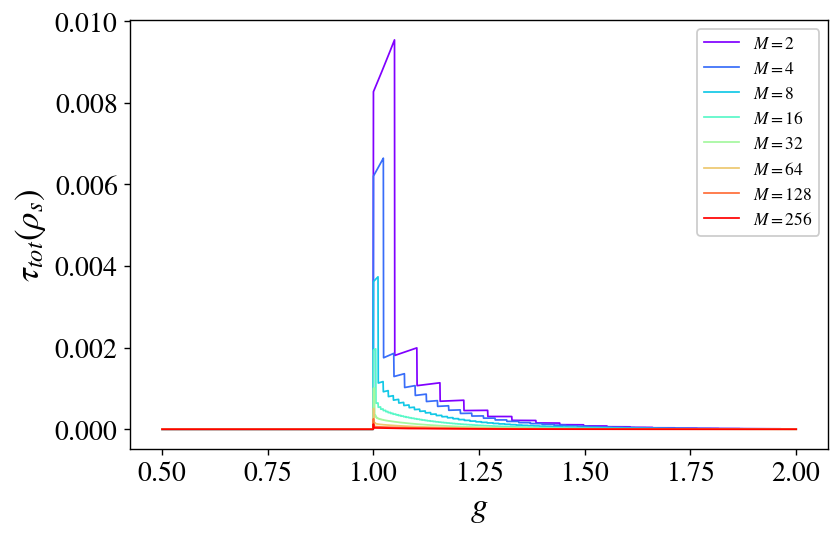}}
    \caption{The figure plots the total two-tangle per qubit versus $g$, derived by numerical calculations, in the cases $M=2,3,4,5,6,7,8,9$. The asymptotic prediction \eqref{qcdapprox}, valid in the limit $k \gg M$, yields uniformly null total two tangle, for all values of $g$.}
    \label{Conc}
\end{figure}

Thus, the entanglement of $\rho_s$, as defined in Eq. \eqref{convex_roof}, is bounded as follows
\begin{equation}
\tau_{tot}(\rho_s)\leq E(\rho_s)\leq \tilde{C}(\rho_s)\, .
\end{equation}
The Dicke states are symmetric by permutation. Thus, $\tau_{tot}(\rho_s)=(M-1){\cal C}_{\mu\nu}^2(\rho_s),\,\forall\mu\neq\nu$, making computations much easier. 
As shown in Figure \ref{Conc}, we found that the total pairwise entanglement, as measured by the two-tangle per qubit, vanishes for large $g$ for any system size. It is furthermore a decreasing function of $M$, thus taking its maximal value for $M=2$, $g\gtrsim1$. It results that, if the entanglement does not vanish for large $M$ or large $g$, it must be multipartite in nature.

The behaviour of the rescaled QCD, shown in Fig. \ref{QCDresc_TC}, stems as a clue suggesting that entanglement isn't asymptotically vanishing in large systems. It might attains a finite value, with a prominent peak in the region $g\gtrsim1$. Furthermore, as previously emphasized, it would be, especially in parametric regions of larger $g$, genuinely multipartite, rather than a mere pairwise entanglement. 

In favour of the occurrence of genuine multipartite entanglement in the ground state of the TC model, let us draw another argument. It was shown that the maximal concurrence of a pure Dicke state is $C(|D^M_{1}\rangle)=2/N$ \cite{koashi_entangled_2000}. It follows the total two-tangle, which equates the ED in this case: $\tau_{tot}(|D^M_{1}\rangle\langle D^M_{1}|)=E(|D^M_{1}\rangle)=4(M-1)/M^2$. Yet, for $M$ even, the Dicke state with $M/2$ excitations is maximally entangled, namely $E(|D^M_{M/2}\rangle = 1$, while its total two-tangle is null. This emphasizes that the entanglement of high-dimensional Dicke states is in general poorly bounded by the quantity $\tau_{tot}$. Its ``entanglement content'' is thus mainly of higher-order, rather than pairwise, and can be extensive. The same observation may be applied to mixed Dicke states, hence our argument in favour of a non-vanishing genuinely multipartite entanglement, in the ground state of the TC model.

In the light of the results presented in this section, entanglement stems as an order parameter, at least for the \textit{finite-sized} superradiant phase transition undergone by the Tavis-Cummings model. It is possibly also the case in the thermodynamic limit, for the considered regime $\eta=10$. Furthermore, it is quite clear that the shape of the distribution \eqref{p_n_asymp}, describing the ground state in the asymptotic regime, strongly depends on the parameters $\eta$ and $g$. It results that an appropriate tuning the latter parameters can lead to a quasi-pure state, and consequently to a non negligible amount of entanglement.

\section{Conclusions} \label{sec:conclusions}

The Tavis-Cummings model undergoes a superradiant QPT that we have shown to be embodied in a SSB mechanism. It corresponds to an infinite sequence of energy level crossings. 
Our study confirms that there exists a critical value $g_c=1$, beyond which the $T$-symmetry introduced above is broken and the cavity field is macroscopically occupied.

We also have answered the question if quantum correlations and entanglement are relevant features in this phenomenon. In fact, the mentioned QPT is associated with a crossover of the quantum correlations of the $M$-qubits density matrix of the system ground state: they are vanishing below the transition point $g_c$, and take finite values above. Quantum correlations between atoms, as measured by the QCD, can thus be considered an order parameter for the superradiant transition in the TC model. Furthermore, this holds both for finite sized systems, as those studied in this work, and in the thermodynamic limit, since the QCD visibly converges towards a finite value, as the system size $M$ increases.

Then, resorting to an entanglement criterion for mixed Dicke states, we showed that this transition is furthermore accompanied by a breaking of the ground state separability. 
By computing the total two-tangle in the system, we were able to employ the monogamy theorem to bound the total entanglement of the system from below. Meanwhile, our previous results pertaining to quantum correlations immediately provided an upper bound.
These results all put together tend to indicate that the system holds a macroscopic amount of entanglement in the broken symmetry phase, and that this entanglement must be multipartite in essence. The precise amount and nature of multipartite entanglement then depends on the values of both the parameters $\eta$ and $g$.

It would thus be of great interest to extend our study to the full phase portrait of this model, namely its behaviour with respect to both parameters. 

We expect our results to be valid in general in superradiant phase transitions; further investigations (i.e. on the Dicke model, etc...) are thus desirable. 

A further interesting subject is to investigate the deep relation between quantum-phase transitions and entanglement in general, and we think that the general method we have proposed here is suitable for this purpose. Our methods could also be profitably applied also in the study of superradiance at finite temperature, where thermal decoherence effects take place.
\section*{Acknowledgements} \label{sec:acknowledgements}
We acknowledge support from the RESEARCH SUPPORT PLAN 2022 - Call for applications for funding allocation to research projects curiosity driven (F CUR) - Project ”Entanglement Protection of Qubits’ Dynamics in a Cavity”– EPQDC and the support by the Italian National Group of Mathematical Physics (GNFM-INdAM). R. F. and A. V. would like to acknowledge INFN Pisa for the financial support to this activity. 

\bibliography{references}

\end{document}